\documentclass[11pt]{article}
\usepackage{graphicx}

\title{A metric for software vulnerabilities classification}
\author{Gabriele Modena}
\date{}
\begin{document}

\maketitle
\begin{abstract}
Vulnerability discovery and exploits detection are two wide areas of
study in software engineering. This preliminary work tries to combine
existing methods with machine learning techniques to define a metric
classification of vulnerable computer programs. First a feature set has
been defined and later two models have been tested against real world
vulnerabilities. A relation between the classifier choice and the features
has also been outlined.
\end{abstract}



\section{Introduction}
A software bug is an error, flaw, mistake, failure, or fault in a computer program that prevents it from behaving as intended. Most bugs arise from
mistakes and errors made by people in either a program’s source code or its design. An unexpected behavior may present several consequences; a
system crash, data loss or corruption or, in other cases, the possibility for a third party to run arbitrary code. The last case is commonly referred
to as vulnerability. Vulnerability exploitation is a core subject of study of computer security; differnt domains present differnt sources of exploitation that lead to a variety of attacks. Software Engineering provides several techniques to discover ﬂaws in code and possible attacks. Many of these techniques work by looking for patterns in source code and or by requiring a user to manually create signatures for exploits, given known attack, to later perform classiﬁcation and recognition. Not every flaw, though, may be an actual vulnerability and signature based recognition often is unable to recognize unseen attacks \cite{Patton01anachilles}. The goal of this paper is to combine software engineering with machine learning techniques to deﬁne a metric to recognize vulnerabilities and to build a model able to classify vulnerable code.

Exploitation of a program can result in severe consequences. Possible goals of a malicious attacker include: gain privileges on a system (root shell on a unix machine); access protected and conﬁdential data (credit card frauds); identity theft (phishing web sites). \cite{ransbotham:exploitation}
Given a programming language, application domain or operating system; vulnerabilities and exploits will present different characteristics and
problematics. This work is focused on the study of a well known class of vulnerabilities found in C/C++ source code:
\begin{itemize}
\item \textbf{Buﬀer and heap overﬂows}: an anomalous condition where a process attempts to store data beyond the boundaries of a ﬁxed-length
buﬀer (or heap). The result is that the extra data overwrites adjacent
memory locations;
\item \textbf{Stack overﬂows}: which occurs when a program writes to a memory address on the program’s call stack outside of the intended data
structure;
\item \textbf{Format string bugs}: attacks that can be used to crash a program or to execute harmful code. The problem stems from the use of unﬁltered
user input as the format string parameter in certain C functions that perform formatting, such as printf()
\end{itemize}
Despite affecting different data structures these vulnerabilities may result in injection into memory and execution of arbitrary code that may result in
an attacker exploiting the system. A typical C/C++ exploit is a program which first aims to fill a memory buffer, place arbitrary assembly code (the shellcode) in memory and take control over the program counter CPU register to ﬁnally execute the injected instructions. Software engineering and computer security provide a set of tools and techniques to both search for ﬂaws in software and detect exploitation attempts on a system.

Two approaches to flaws discovery in software can be identified: \textit{static analysis} consists of scanning the source code of a program to detect patterns
of known errors \cite{Naumovich:1997:ASA:267895.267904};  \textit{dynamic analysis} consists of executing a computer program and monitor its behavior on the host system (memory tracing, disk access, cpu usage) \cite{Allen98specifyingand}.
The \textit{preliminary work} presented in this paper is focused on static analysis; this approach presents two main drawbacks. First static analysis tools can not accurately discriminate between a general flaw and a vulnerability. Second, many bugs can not be detected at all. In order to address this issue we model vulnerability identification as a  classification task. We identified a set of feature that characterize the state of a computer program. We collected a dataset of real world vulnerable code bases and we annotated it according to the identified features. Finally we train and tested two classifiers, Support Vector Machines and
Fisher Discriminant Analysis, to identify vulnerable programs. 

The paper is organized as follows. In Section 2 we describe a model to classify vulnerabilities and exploits and we introduce a taxonomy of feature to characterize code bases. In Section 3 we report experimental results obtained on a dataset we collected. Section 4 conclused the paper.

\section{A model to classify vulnerabilities and exploits}
\subsection{Exploit classiﬁcation}
As a first step in our work we looked at correlations between a vulnerability and its exploit that could provide a metric able to generate dynamic signatures.
Such a technique has been used with success in polymorphic worm detection \cite{YangSRCZM06}. However, worms seem to be easier to characterize than this particular kind of attacks. From the point of view of the code exploits are all quite similar to each other and present an extremely simple structure which is difficult to extract information from. Computational Linguistic has been proved to obtain good results in exploit detection \cite{RieckL06}. Given time constraints it was not possible to explore this approach in detail. In further work it could be interesting to study how this approach can be combined to machine learning to reach an even higher accuracy.
Similarly to \cite{ChallagullaBYP05} and \cite{Shin:2011:ISU:1988630.1988632} the focus of our work is on feature engineering and the definition of a metric to evaluate code quality in relationship to possibly exploitable flaws. We identify a taxonomy of features at different aggregation levels and we consider both quantitative and qualitative aspects. As a point of departure we defined features and complexity metrices based not on code properties only. We take into account social indicators as parts of the software development process.

\subsection{Vulnerability classiﬁcation}
Despite the above mentioned problems, machine learning may still be used to discriminate vulnerable source code from good one. As a simpliﬁcation
ﬂawed but not vulnerable code is still regarded as good in this model. This approach may seem strict, but it is important to note that software engineering already allows us to ﬁnd ﬂaws in sources. The further step this model aims to reach is to determine which of these ﬂaws may result in possible exploitable vulnerabilities. The most important task in building the model was the choice of the correct features to properly characterize a vulnerability. In order to deﬁne a proper feature set, information extracted using code auditing tools has been enriched by metadata discovered by analyzing the environment in which the software has been developed. To test the model, a dataset has been collected by manually annontating existing vulnerable and exploited open source software. 

We used Fisher Discriminant Analysis (FDA) and Support Vector Machines (SVM) to perform classification. Further experiments have been performed by testing against a subset of features selected using Recursive Feature Elimination (RFE). The experiments have been carried out using the \textit{mlpy} suite \footnote[4]{http://mlpy.fbk.eu}.

\subsection{Dataset}
The dataset consists in 75 instances of ﬂawed open source software. 50 of these code ﬂaws are vulnerabilities for which an exploit has been developed. 25 instances are ﬂaws that resulted in misfunctioning of the program without introducing security risks. It is worth noticing that from a syntactic point of view regular ﬂaws and vulnerabilities may often look very similar. What makes a flaw exploitable is the type of data structure it aﬀects and its relationship with other sensible parts of the code.
The vulnerable data has been collected from the 2008 GLSA vulnerability list \footnote[3]{http://www.gentoo.org/security/en/glsa}. Buggy code samples have been collected from projects bug repositories. Instances have been manually labelled and annotated according to the previously described feature set.

Three layers of features have been deﬁned.
\begin{itemize}
\item Features extracted by performing static analysis (layer 1 )
\begin{itemize}
\item write operations on buﬀers (strings, integers)
\item NULL pointer dereferences
\item access to storage that may have been deallocated
\item memory leaks
\item returning a pointer to stack-allocated storage
\item the value of a location is used before it is deﬁned
\end{itemize}
These features have been selected by performing static analysis on
source code using splint \footnote[1]{http://www.splint.org} and cppcheck \footnote[2]{http://sourceforce.net/projects/cppcheck/}. Given the nature of vulnerabilities we want to classify flaws that may result in memory tampering will help to classify the software as possibly dangerous.

\item Features extracted by parsing flawed code
\begin{itemize}
\item use of safe libraries (buﬀer checking)
\item presence and complexity of if-then-else statements (layer 2 )
\item presence of loops
\item nature of the software (server vs. client application)
\item number of calls to [cm]alloc
\item size of the code
\item use of recursive functions (may result in memory leaks). We run a parsing over vulnerable code to characterize the environment surrounding the features of layer 1. Use of safe libraries may prevent memory tampering. A procedure with an high number of loops, recursive functions or complex if-then-else statements, on the
other end, may be prone to human logical mistakes. Also the nature of the software should be taken into account, since a server application may be implemented using different routines and design than a client.
\end{itemize}
\item Metadata extracted from bugzilla, mailing lists, statistics and charts (layer 3 )
\begin{itemize}
\item age of the code (more recent code is more likely to contain flaws)
\item number of committers
\item popularity of the program (the most popular, the more it has been analyzed)
\item popularity of the platform it is running on
\item kind of platform
\item reputation of the developers
\item relation to security applications
\item status of the code
\item legacy vs. development
\item exploitation history
\end{itemize}
\end{itemize}
These properties describe a social component behind the development of the software. By analyzing real world vulnerabilities it is possible to try to characterize the environment in which mistakes have been committed. The age of the code and the number of people involved may give useful information. The popularity of the program, operating system and architecture on which the software runs may be indexes of the eﬀort spent in debugging, testing and peer reviewing. Teams of
developers with high reputation in providing tested and clean code or the relation of a program to security can be used to judge the risk of
a ﬂaw as well. The status of the code may give information about the current development and maintenance of the project.
These information have been extracted by checking bugzilla repositories, charts of the number of ﬂaws for a given program, quality statistics derived from comments and threads on mailing lists and newsgroups.
Features of layer 1 and layer 2 describe respectively the nature of ﬂaws found in code and properties of the software. This is the kind of information
usually searched for by security analysts to later test for an actual vulnerability by developing a proof of concept. The role of layer 3 is to enrich
these properties by weighting the possible risk of a software flaw being a vulnerability given the status of the project and its history..

\section{Results}

\begin{table}
\centering
\begin{tabular}{ l | c | c | r }
  \hline \hline
   & L1+L2 & all features & RFE \\ \hline
  FDA & .38 & .58 & .63  \\ \hline
  SVM & .41 & .63 & .71 \\ \hline
\end{tabular}
	\caption{Vulnerability classification accuracy (percentage)}
\end{table}

\textit{Table 1} summarizes the results obtained with weighted SVMs and FDA on different features subsets. Weights have been assigned based on the
presence of features from layer 3. The tests have been performed using 75\% of the dataset for training and the remaining 25\% as test set. In order
to prevent overﬁtting 10 fold cross-validation and bootstrapping have been used to perform evaluation. Anyway, no sensible difference in accuracy has
been noticed performing tests without cross-validation and bootstrapping. 
Reasons for this result can be justiﬁed by the small size of the dataset and by the fact that learners with a strong linear bias has been used.
The ﬁrst experiment, \textit{L1+L2}, uses only the ﬁrst two layers for training and classiﬁcation, thus resulting in an unweighted classiﬁers that resembles
the behavior of a human researching for vulnerabilities. As expected, the results in terms of accuracy are close to random choice. Given the presence
of a flaw, it is difficult to classify it as a vulnerability without actually coding a proof of concept.

The second experiment, \textit{all features}, uses all three layers. We report an improvement in terms of accuracy that seems to conﬁrm the importance of
the social component in classifying the vulnerabilities as such. Given that a source is ﬂawed, layer 3 provides useful information
Finally the third experiment has been carried on by selecting features via RFE. By taking into account only the most
relevant features we reported a performance boost in the accuracy of both classifiers. 

The results that emerge from this experiments seem to conﬁrm that using machine learning and taking into account factors not directly related
to the code can improve the classiﬁcation accuracy; moreover, despite a difference in terms of performance, the classiﬁer does not seem to play a big
role compared to the feature set. It is interesting to note that RFE deleted most of the features of layer 1 and 2, where layer 3 showed to carry much information. In particular determinant features seem to be the presence and complexity of if-then-else statements, popularity of the program and platform, the relation with security
applications.

\section{Conclusion}
The aim of this project was to combine software engineering with machine learning techniques to achieve vulnerabilities classiﬁcation and exploit classi-
ﬁcation. While it was not possible to recognize attacks, detecting vulnerable code could beneﬁt from machine learning. A result that emerge from this
study is that the feature set seem to have a dominant role over the classiffication model.

Further improvements on this study would be extending the dataset with more instances and add or remove features given the results highlighted by
RFE. Moreover, manually labeling the instances may have introduced bias. A large scale survey on security mailing lists or open source teams may lead
to a more reliable and complete dataset. In terms of exploit recognition, it could be interesting to ﬁnd a method to combine this model with language models in order to build a dynamic signature generator.

\bibliographystyle{plain}
\bibliography{lib}

\end{document}